\documentclass[prb]{revtex4}% Physical Review B

\usepackage{graphicx}% Include figure files
\usepackage{dcolumn}% Align table columns on decimal point
\usepackage{bm}% bold math

\begin{document}

\preprint{APS/123-QED}

\title{Comment on `A one-way speed of light experiment'}% Force line breaks with \\

\author{Rodrigo de Abreu}%
\affiliation{%
Departamento de F\'{\i}sica, Instituto Superior T\'{ecnico}, 1049-001 Lisboa, Portugal
}%
%Lines break automatically or can be forced with \\
\author{Vasco Guerra}
 \email{vguerra@alfa.ist.utl.pt}
% \affiliation{Departamento de F\'{\i}sica}
 \altaffiliation[Also at ]{Centro de F\'{\i}sica dos Plasmas, I. S. T., 
1049-001 
Lisboa, Portugal}

\date{\today}% It is always \today, today,
             %  but any date may be explicitly specified

\begin{abstract}

A recent paper published in \textit{Am. J. Phys.} describes an experiment designed to measure the one-way speed of light.\cite{Greetal2009}
Although the experiment is very interesting, in particular to be used in student laboratories, 
it is in fact determining the two-way speed of light.

\end{abstract}

% PACS, the Physics and Astronomy
                             % Classification Scheme.
%Use showkeys class option if keyword
                              %display desired
\maketitle

%\section{Introduction}

Greaves, Rodr\'{\i}guez and Ruiz-Camacho have recently published a very interesting paper in this journal, entitled ``A one-way speed of 
light experiment''.\cite{Greetal2009} The authors correctly refer to the literature when stating that the problem of clock synchronization
has been used to show that the one-way speed of light is a quantity that cannot be measured. They further observe that ``any measurement
of the speed of light in one direction, from A to B, which uses only one clock to avoid the problem of synchronization, requires the 
return back to A of the time information of the arrival at B''. 
The experiment they propose is an attempt to avoid this situation. Nevertheless, it occurs as well. 

As pointed out by
Finkelstein,\cite{Finkelstein2009} what the authors indeed measure is the two-way speed of light, if the coaxial cable of length $L$ is equivalent to a 
light signal propagating in vacuum with speed $c$, as the fixed time delay is assumed to be 
$\Delta t=L/c = 23.73\ \mbox{m}/3\times10^8\ \mbox{m/s}=79$ ns.
By using the value $c$ to calculate $\Delta t$, 
in fact the authors have implicitly used two clocks, one at the photosensor and one at the oscilloscope, 
``synchronized'' according to Einstein's procedure.

The experiment described corresponds to the example of ``synchronizing'' clocks with the speed of a $F_1$ car presented in a
previous paper.\cite{AG2008}
It is impossible to measure the speed of a $F_1$ between two points in a circuit by using its average speed along the track 
to ``synchronize'' clocks. Of course the $F_1$ car
has a speed, but this speed is not the difference of times at the two positions if the clocks have been previously set using the average speed of the car.
To make the analogy precise, consider a track of length $L$ and let the $F_1$ car do several laps along the circuit, always in the same way (braking on
the same positions, accelerating on the same positions, etc.). With one clock at the start-finish line, $C_0$, we can determine the average speed of the
$F_1$ on the circuit, $\bar{v}$, by measuring the time it takes to complete one lap. 
Now, if we have another clock, $C_1$, on a position of the circuit at a distance $L_1$ from the start-finish line, we can
``synchronize'' this clock with the one at the origin by setting $C_0$ to zero when the car passes at the
start-finish line and by setting $C_1$ to $L_1/\bar{v}$ when the car passes at in front of $C_1$ 
(note that this is exactly what it is done in the paper by Greaves \textit{et al} \cite{Greetal2009} by
assuming a constant time delay of 79 ns).
Of course we can now ``measure'' the ``speed'' of the car
on the remaining portion of the circuit, corresponding to a distance $L-L_1$. With the clocks set in the way just described, we will conclude, 
without surprise, that this ``speed'' is $\bar{v}$, regardless of the value of $L_1$. 
Similarily, the measurement by Greaves \textit{et al} \cite{Greetal2009} for the ``speed'' of 
light gives $c$.

What has been measured is not the one-way speed of light, it is the two-way speed of light or, equivalently,
the one-way ``Einstein speed'' of light, formerly defined.\cite{GA2006b} 
Of course this speed is rigorously $c$, because 
the procedure has been done in such a way that it cannot be otherwise!
There is an implicit synchronization in Greaves \textit{et al},\cite{Greetal2009} where the clocks have been ``synchronized'' operationally.
There is no problem in doing so, but we cannot attribute physical meaning to a perfectly defined quantity (the Einstein speed)
other than its true meaning. If we do not know the one-way speed of light in one frame, we cannot use $c$ to cope with half of the circuit (the delay
on the coaxial cable) and then pretend we have measured the one-way speed of light in the remaining part of the circuit.
The knowledge of the one-way ``Einstein speed'' of light is just a result of a definition, based on an experimental fact: the constancy
of the \textit{two-way} speed of light. So far, this knowledge does not justify the assumption of the constancy of the one-way speed
of light in all frames, although we can work, for operational reasons, \textit{as if} the one-way speed of light was $c$.\cite{GA2006b,AG2008}

\bibliographystyle{unsrt}

\end{document}